\documentclass[aps,axodraw,twocolumn,prl]{revtex4}
\usepackage{fancyhdr}
\usepackage{fancybox}
\usepackage{graphicx}
\usepackage{color}
\usepackage{soul}
\usepackage{latexsym}
\usepackage{graphicx}
\usepackage{amsmath}
\usepackage{amsfonts}
\usepackage{amssymb}
\usepackage{feynmp}
\def\beq{\begin{equation}}
\def\eeq{\end{equation}}
\def\bea{\begin{eqnarray}}
\def\eea{\end{eqnarray}}
\def\ie{{\it i.e.}}

\def\lsim{\mathrel{\raise.3ex\hbox{$<$\kern-.75em\lower1ex\hbox{$\sim$}}}}
\def\gsim{\mathrel{\raise.3ex\hbox{$>$\kern-.75em\lower1ex\hbox{$\sim$}}}}
\def\ifmath#1{\relax\ifmmode #1\else $#1$\fi}

\def\fbi{~{\mbox{fb}^{-1}}}

\def\gev{~{\mbox{GeV}}}

\def\to{\rightarrow}

\def\wtil{\widetilde}

\def\anti{\overline}

    \def\fillboxx#1#2{\hbox to #1{\vbox to #2{\vfil}\hfil}   }

\def\gev{~{\rm GeV}}

\def\cnone{\wt\chi^0_1}

\def\cntwo{\wt\chi^0_2}

\def\smu{\wt\mu}

\def\wt{\widetilde}
\def\anti{\overline}

\def\stau{\wt \tau}

\def\gl{\wt g}

\def\sq{\wt q}

\def\slep{\wt \ell}

\newcommand{ \slashchar }[1]{\setbox0=\hbox{$#1$}   
   \dimen0=\wd0                                     
   \setbox1=\hbox{/} \dimen1=\wd1                   
   \ifdim\dimen0>\dimen1                            
      \rlap{\hbox to \dimen0{\hfil/\hfil}}          
      #1                                            
   \else                                            
      \rlap{\hbox to \dimen1{\hfil$#1$\hfil}}       
      /                                             
   \fi}     

\begin{document}
\title{Accurate Mass Determinations in Decay Chains with Missing Energy}

\author{Hsin-Chia Cheng${}^{a}$, Dalit Engelhardt${}^{b}$, John
  F. Gunion${}^{a}$, Zhenyu Han${}^{a}$,  and Bob McElrath${}^{c}$}
\address{ ${}^{a}$Department of Physics, University of California, Davis, CA 95616,\\
${}^{b}$Department of Physics, Boston University, Boston, MA 02215,
${}^{c}$CERN, Geneva 23, Switzerland}

\begin{abstract}
Many beyond the Standard Model theories include a stable dark matter candidate that
yields missing / invisible energy in collider detectors. If observed at the Large
Hadron Collider, we must determine if its mass and other
properties (and those of its partners) 
predict the correct dark matter relic density.  We give a new procedure
for determining its mass with small error.
\end{abstract}
\maketitle

\begin{figure}[h]
\begin{center}
 \includegraphics[width=0.27\textwidth]{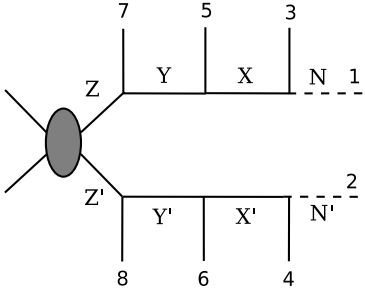}
\vspace*{-.1in}
\caption{\label{fig:topology} The event topology.}
\vspace*{-.3in}
\end{center}
\end{figure}

One of the most dramatic possibilities for the Large Hadron Collider
(LHC) is observation of events with large missing energy compatible
with the production of a stable, weakly-interacting particle that
could explain the universe's relic dark matter content. Many beyond
the Standard Model (SM) theories contain such a particle, denoted
$N$.  In particular, in the Minimal Supersymmetric Standard Model (MSSM) the
lightest neutralino $\cnone$ is stable if $R$-parity is conserved.
Each LHC event must contain two $N$'s that each emerge at the end
of a chain decay. For example, in the MSSM, a large
production rate is associated with squark pair, $\wtil q \wtil q$,
production, and each $\wtil q$ can have substantial
probability to decay via $\wtil q\to q\cntwo\to q \slep \ell \to q \ell\anti\ell
\cnone$ ($\ell=e,\mu,\tau$), where $\cntwo$ and $\wtil l$ are the 2nd lightest neutralino
and slepton, respectively. More generally, we will use the notation
$Z\to 7+Y \to 7+5+X \to 7+5+3+1(=N)$, where particles 7, 5 and 3 are Standard Model 
jets or leptons and $Z$, $Y$, and $X$ are the intermediate on-shell 
resonances of the model in
question. This event
structure is illustrated in Fig.~\ref{fig:topology}.  This letter
gives a procedure for accurately determining $M_Z$, $M_Y$,
$M_X$ and $M_N$ for this topology.

Many mass determination procedures in the literature examine only one
decay chain at a
time~\cite{Allanach:2000kt,Kawagoe:2004rz,Gjelsten:2004ki,Miller:2005zp}.
This often does not allow one to solve for the event's missing
momenta. An exception is a very long decay chain starting from the
gluino, as discussed in Ref.~\cite{Kawagoe:2004rz}. However, in the
actual analysis the $\cntwo$, $\slep$ and $\cnone$ masses were assumed
to be known and only the gluino and sbottom masses were
fitted~\cite{Allanach:2004ub}. Considering both decay chains
simultaneously can potentially give us more information and allow a
better determination of the
masses~\cite{Cheng:2007xv,Cho:2007qv,Nojiri:2007pq}.  Our current
procedure does this for the decay chains of Fig.~\ref{fig:topology}.
If all particles can be correctly located on the decay chains and
there are no experimental effects, then by considering two events we
can solve for {\it all} the 4-momenta in both events and determine all
the masses up to a discrete ambiguity. After examining a small number
of event pairings, a unique solution will emerge.

Assuming we can isolate LHC events with the topology in Fig.~\ref{fig:topology}
 and using $m_N=m_{N'}$, $m_X=m_{X'}$, $m_Y=m_{Y'}$, $m_Z=m_{Z'}$, we have the following constraints, 
\begin{equation}
\label{massshell}
\begin{array}{lcrcl}
(M_Z^2 &=)& (p_1+p_3+p_5+p_7)^2&=&(p_2+p_4+p_6+p_8)^2,\\
( M_Y^2 &=)& (p_1+p_3+p_5)^2&=&(p_2+p_4+p_6)^2,\\
(M_X^2 &=)& (p_1+p_3)^2&=&(p_2+p_4)^2,\\
(M_N^2 &=)& p_1^2&=&p_2^2.
\end{array}
\end{equation}
where $p_i$ is the 4-momentum for particle $i$ $(i=1\ldots 8)$.
Since the only invisible particles are $1$ and $2$ and since we can
measure the missing transverse energy,
there are two more constraints:
\begin{eqnarray}
p_1^x+p_2^x=p_{miss}^x,\quad 
p_1^y+p_2^y=p_{miss}^y.\label{misspt}
\end{eqnarray}
Given the 6 constraints in Eqs.~(\ref{massshell}) and (\ref{misspt}) 
and 8 unknowns from the 4-momenta of the missing particles, there remain two 
unknowns per event.  The system is under-constrained and cannot be
solved.
This situation changes if we use a second event with the same decay
chains, under the assumption that the invariant masses are the same in the two
events. Denoting the 4-momenta in the second event as $q_i$
$(i=1\ldots 8)$, we have 8 more unknowns, $q_1$ and $q_2$, but 10 more
equations,
\begin{eqnarray}
&&\begin{array}{rcccl}
q_1^2&=&q_2^2&=&p_2^2,\\
(q_1+q_3)^2&=&(q_2+q_4)^2&=&(p_2+p_4)^2,\\
(q_1+q_3+q_5)^2&=&(q_2+q_4+q_6)^2&=&(p_2+p_4+p_6)^2,\\
\multicolumn{5}{c}{
\begin{array}{rcl}
(q_1+q_3+q_5+q_7)^2&=&(q_2+q_4+q_6+q_8)^2\\
    &=&(p_2+p_4+p_6+p_8)^2,\\
\end{array}
} 
\end{array}
\label{misspt2}
\nonumber \\ 
&&q_1^x+q_2^x=q_{miss}^x,\quad
q_1^y+q_2^y=q_{miss}^y.
\end{eqnarray} 
Altogether, we have 16 unknowns and 16 equations. The system can be
solved numerically and we obtain discrete solutions for $p_1$, $p_2$,
$q_1$, $q_2$ and thus the masses $m_N$, $m_X$, $m_Y$, and $m_Z$. Note
that the equations always have 8 complex solutions, but we will keep
only the real and positive ones which we henceforth call ``solutions''.
Further details regarding practical and high-speed techniques for
obtaining the solutions will appear in a future paper~\cite{massesii}.

For illustration and easy comparison to the literature, we apply our method for the SUSY point, SPS1a \cite{Allanach:2002nj},
although many of the discussions below apply for generic cases. For
SPS1a, the particles corresponding to $N,X,Y,Z$ are $\cnone$,
$\slep_R(\ell=e/\mu)$, $\cntwo$, $\sq_L(q=d,u,s,c)$
respectively.  The
masses are \{97.4, 142.5, 180.3, 564.8/570.8\} GeV, with the final two
numbers corresponding to up/down type squarks respectively.  Since $m_{\wtil \tau}\neq m_{\wtil e,\wtil\mu}$, the
$\ell=\tau$ case is an important background. We generate events with PYTHIA 6.4 \cite{Sjostrand:2006za}.

\begin{figure}
\begin{center}
 \includegraphics[width=0.4\textwidth]{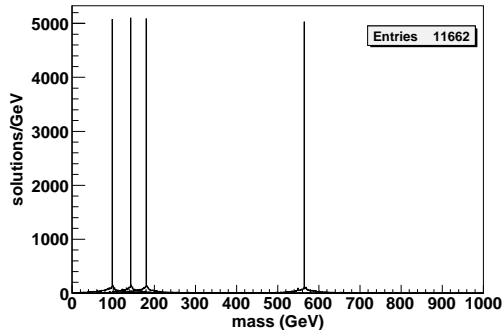}
\vspace*{-.1in}
\caption{\label{fig:ideal}We plot the number of mass solutions (in 1
  GeV bins --- the same binning is used for the other plots) vs. mass in the ideal case. All possible pairs for 100 events are included.}
\vspace*{-.3in}
\end{center}
\end{figure}

We first consider the ideal case: no background events, all
visible momenta measured exactly, all intermediate particles 
on-shell and each visible particle associated with the correct decay
chain and  position in the decay chain. We also restrict the squarks
to be up-type only. In this case, we can solve for the masses exactly
by pairing any two events. The only complication comes from there being 
8 complex solutions for the system of equations,
of which more than one can be real and positive. Of course, the
wrong solutions are different from pair to pair, but the correct
solution is common. The mass distributions for the ideal case with 100
events are shown in Fig.~\ref{fig:ideal}. As expected, we observe
$\delta$-function-like mass peaks on top of small backgrounds coming
from wrong solutions. On average, there are  about 2 solutions per pair of events. 

The $\delta$-functions in the mass distributions arise only when exactly
correct momenta are input into the equations we solve.  
To be experimentally realistic, we now include the following.
\begin{figure}
\begin{center}
 \includegraphics[width=0.4\textwidth]{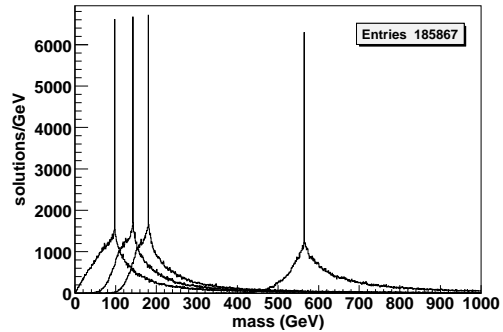}
\vspace*{-.1in}
\caption{\label{fig:allcombis}Number of mass solutions versus mass after
  including all combination pairings for 100 events.}
\vspace*{-.3in}
\end{center}
\end{figure}

1. {\bf Wrong combinations.} For a given event a
``combination'' is a particular assignment 
of the jets and leptons to the external legs of Fig.\ref{fig:topology}.
For each event, there is
only one correct combination (excluding $1357
\leftrightarrow 2468$ symmetry).  Assuming that we can
identify the two jets that correspond to the two quarks, we have 8
(16) possible combinations for the $2\mu2e$ ($4\mu$ or $4e$) channel.
The total number of combinations for a pair of events is the product
of the two, \ie\ 64, 128 or 256. Adding the wrong combination pairings for the
ideal case yields the mass distributions of Fig.~\ref{fig:allcombis}.
Compared to Fig.~\ref{fig:ideal}, there are 16 times more (wrong)
solutions, but the $\delta$-function-like mass peaks remain evident.

2. {\bf Finite widths.} For SPS1a, the widths of the intermediate particles are roughly
5~GeV, 20~MeV and 200~MeV for $\sq_L$, $\cntwo$ and $\slep_R$.  Thus,
the widths are quite small in comparison to the corresponding masses.

3. {\bf Mass splitting between flavors.} The masses for up and down type
squarks have a small difference of 6 GeV. Since it is impossible to
determine flavors for the light jets, the mass determined should be
viewed as the average value of the two squarks (weighted by the parton
distribution functions).

4. {\bf Initial/final state radiation.} These two types of radiation not only smear the visible particles' momenta, but also provide a source for extra jets in the events.  We will apply a $p_T$ cut to get rid of soft jets.  

5. {\bf Extra hard particles in the signal events.} In SPS1a, many of
the squarks come from gluino decay ($\gl\rightarrow q\sq_L$), which yields
another hard $q$ in the event. Fortunately, for SPS1a 
$m_{\gl} - m_{\sq_L}=40\gev$  is much
smaller than $m_{\sq_L}-m_{\cntwo}=380\gev$.
Therefore, the $q$ from squark decay is usually much more energetic
than the $q$ from $\gl$ decay.  We select the two jets with highest
$p_T$ in each event after cuts.  Experimentally one would want to
justify this choice by examining the jet multiplicity to ensure that
this analysis is dominated by 2-jet events, and not 3 or 4 jet events.
Furthermore, the softer jets will be an indication of clearly separable
mass-differences.

6. {\bf Background events.}  The SM backgrounds are negligible
for this signal in SPS1a. There are a few significant backgrounds from other SUSY processes:

(a) $\sq_L\rightarrow q\cntwo\rightarrow q\tau\stau\rightarrow
  q\tau\tau\cnone$ for one or both decay chains, with all $\tau$'s
  decaying leptonically. Indeed,  $\cntwo\rightarrow\tau\stau$ has the
  largest partial width, being 14 times that of
  $\cntwo\rightarrow\mu\smu$.  However, to be included in our
  selection the two $\tau$'s
  in one decay chain must both decay to leptons with the same flavor, which
  reduces the ratio. A cut on lepton $p_T$ also
  helps to reduce this background, since leptons from $\tau$ decays are
  softer.  Experimentally one should perform a separate search for
  hadronically decaying tau's or non-identical-flavor lepton decay chains to
  explicitly measure this background.
  
  (b) Processes containing a pair of sbottoms, especially $\wt b_1$.
  In SPS1a the first two generations of squarks are nearly degenerate.
  In any model, they must be discovered in a combined analysis since
  light quark jets are not distinguishable.  Well-separated squark
  masses would show up as a double peak structure in $M_Z$.  However
  $b$ jets are distinguishable and a separate analysis should be
  performed to determine the $b$ squark masses.  This presents a
  background to the light squark search since $b$-tagging efficiency is
  only about 50\% at high $p_T$.

(c) Processes that contain a pair of $\cntwo$'s, not both
  coming from squark decays. For these events to fake signal events,
  extra jets need to come from 
  initial and/or final state radiation or other particle decays. For
  example, direct $\cntwo$ pair production or $\cntwo+\gl$
  production. These are electroweak processes, but, since $\cntwo$ has a
  much smaller mass than squarks, the cross-section is not negligible.
  In our SPS1a analysis, the large jet $p_T$ cut reduces this kind of
  background due to the small $m_{\gl} - m_{\sq_L}$.

7. {\bf Experimental resolutions.} In order to estimate this experimental
effect at the LHC, we process all events with ATLFAST\cite{atlfast}, a fast
simulation package of the ATLAS detector. Since we assume 300
$fb^{-1}$ integrated luminosity, we run ATLFAST in the high
luminosity mode.

The cuts used to isolate the signal are:

I) 4 isolated leptons with $p_T>10$ GeV, $|\eta|<2.5$ and
matching flavors and charges consistent with our assumed $\cntwo \to
\slep \to \cnone$ decay;

II) No $b$-jets and $\geq 2$ jets with $p_T>100$ GeV,
$|\eta|<2.5$. The 2 highest-$p_T$ jets are taken to be particles 7
and 8; 

III) Missing $p_T>50$ GeV.

\begin{figure}
\begin{center}
 \includegraphics[width=0.4\textwidth]{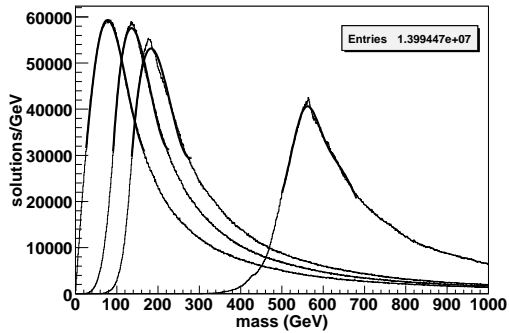}
\vspace*{-.1in}
\caption{\label{fig:smeared}Mass solutions with all effects 1 -- 7
  included and after cuts I
  -- III for the SPS1a SUSY model and $L=300\fbi$.}
\vspace*{-.3in}
\end{center}
\end{figure}

\noindent For a data sample with 300 $fb^{-1}$ integrated luminosity, there are
about 1050 events left after the above cuts, out of which about 700
are signal events.  After taking all possible pairs for all possible
combinations and solving for
the masses, we obtain the mass distributions in
Fig.~\ref{fig:smeared}.

Fitting each distribution using a sum of a Gaussian plus a (single)
quadratic polynomial and taking the maximum positions of the fitted
peaks as the estimated masses yields \{77.8, 135.6, 182.7, 562.0\}$\gev$.
  Averaging over 10
different data samples, we find 
\bea 
&&m_N=76.7\pm1.4\gev,\quad
m_X=135.4\pm1.5\gev,\cr 
&&m_Y=182.2\pm1.8\gev,\quad m_Z=564.4\pm2.5\gev.\nonumber
\eea 
The statistical uncertainties are very small, but
there exist biases, especially for the two light masses. In practice,
we can always correct the biases by comparing real data with Monte
Carlo. Nevertheless, we would like to reduce the biases as much as
possible using data only. In some cases, the biases can be
very large and it is essential
to reduce them before comparing with Monte Carlo.

The combinatorial background is an especially important source of bias
since it yields peaked mass distributions that are
not symmetrically distributed around the true masses,
as can be seen from Fig.~\ref{fig:allcombis}.  This will introduce
biases that survive even after smearing. Therefore, we concentrate on
reducing wrong solutions.

First, we reduce the number of wrong combinations by the following
procedure.  For each combination choice, $c$, for a given event, $i$
($i=1,N_{evt}$), we count the number, $N_{pair}(c,i)$, of events that
can pair with it (for some combination choice for the 2nd events) and
give us solutions. We repeat this for every combination choice for
every event. Neglecting effects 2.-- 7., $N_{pair}(c,i)=N_{evt}-1$ if
$c$ is the correct combination for event $i$. After including
backgrounds and smearing, $N_{pair}(c,i)<N_{evt}-1$, but the correct
combinations still have statistically larger $N_{pair}(c,i)$ than the
wrong combinations.  Therefore, we cut on $N_{pair}(c,i)$.
For the SPS1a model point, if $N_{pair}(c,i)\leq
0.75\,N_{evt}$  we discard the combination
choice, $c$, for event $i$.  If all possible $c$ choices for event $i$
fail this criterion, then we discard event $i$ altogether 
(implying a smaller $N_{evt}$ for the next analysis cycle). We then repeat
the above procedure for the remaining events
until no combinations can be removed. After
this, for the example data sample, the number of events is reduced
from 1050 (697 signal + 353 background) to 734 (539 signal + 195
background), and the average number of combinations per event changes from 11 to 4.

Second, we increase the significance of the true solution by weighting
events by $1/n$ where $n$
is the number of solutions for the corresponding pair (using only the
combination choices that have survived the previous cuts).  This causes
each pair (and therefore each event) to have equal weight in our
histograms.  Without this weighting, a pair with multiple solutions has more
weight than a pair with a single solution, even though at most one
solution would be correct for each pair.

\begin{figure}
\begin{center}
 \includegraphics[width=0.4\textwidth]{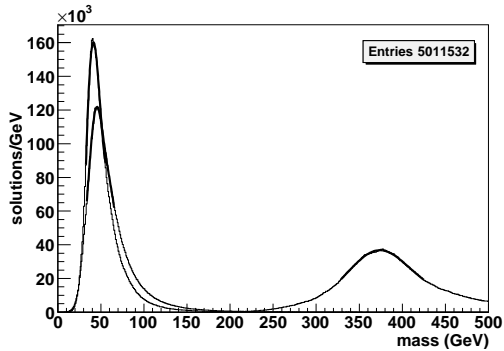}
\vspace*{-.1in}
\caption{\label{fig:dmass}SPS1a, $L=300\fbi$ mass difference distributions. }
\vspace*{-.3in}
\end{center}
\end{figure}

\begin{figure}[t!]
\begin{center}
 \includegraphics[width=0.4\textwidth]{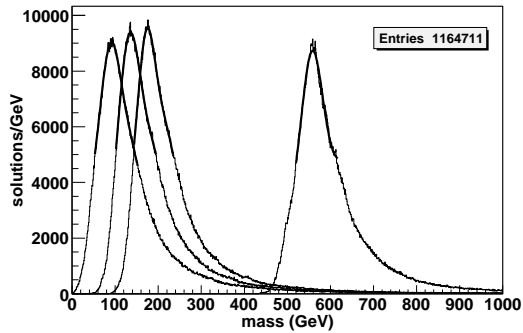}
\vspace*{-.1in}
\caption{\label{fig:masses_final}Final mass distributions after
  the bias-reduction procedure for the SPS1a SUSY model and $L=300\fbi$.}
\vspace*{-.3in}
\end{center}
\end{figure}

Finally, we exploit the fact that wrong solutions and
backgrounds are much less likely to yield $M_N, M_X, M_Y$, and $M_Z$ values that are
all simultaneously close to their true values. We plot the
$1/n$-weighted number of solutions as a function of the three mass
differences (Fig.~\ref{fig:dmass}).  We define mass difference windows
by $0.6\times $ peak height and keep only those solutions
for which {\it all three} mass differences fall within the mass
difference windows.
The surviving solutions are plotted (without the $1/n$ weighting) in
Fig.~\ref{fig:masses_final}. Compared
with Fig.~\ref{fig:smeared}, the mass peaks are narrower, more
symmetric and the fitted values are less biased. The fitted masses
are \{91.7,135.9, 175.7  558.0\} GeV. Repeating the procedure for 10
data sets, we find
\bea
&&m_N=94.1\pm2.8\gev, \quad m_X=138.8\pm2.8\gev,\cr
&&m_Y=179.0\pm3.0\gev,\quad m_Z=561.5\pm4.1\gev.\nonumber
\eea
Thus, the biases are reduced at the cost of (slightly) increased statistical errors. 

We have applied our method to other mass points to show its
reliability.
Details will be presented in \cite{massesii}.  We quote here
results for  ``point 1''  defined in
Ref.~\cite{Cheng:2007xv} with the following masses: \{85.3, 128.4,
246.6, 431.1/438.6\} GeV.  For 100  fb${}^{-1}$ data, we
have about 1220 events (1160 signal events) after the
pre-bias-reduction cuts. After following a bias reduction procedure
and using 10 data samples, we obtain
$m_N=85\pm4\gev$, $m_X=131\pm4\gev$, $m_Y=251\pm4\gev$, $m_Z=444\pm 5\gev$.

We emphasize that the remaining biases in the above mass determinations can
be removed by finding those input masses that yield the observed
output masses after processing Monte Carlo generated data through our
procedures. In this way, very accurate central mass
values are obtained with the indicated statistical errors.

The above results for the $N$, $Y$ and $X$ masses for 
the SPS1a point and point \#1 can be compared to those obtained 
following a very different procedure in
Ref.~\cite{Cheng:2007xv}. There, only the $X\to Y \to N$ parts of the
two decay chains were employed and we used only $4\mu$ events. For the
SPS1a model point we obtained $m_N=98\pm9\gev$, $m_Y=187\pm10\gev$, and
$m_X=151\pm 10\gev$. And, for point \#1 we found $m_N=86.2\pm
4.3\gev$, $m_X=130.4\pm 4.3\gev$ and $m_Y=252.2\pm 4.3\gev$. Including
the $4e$ and $2\mu 2e$ channels will reduce these errors by a factor
of $\sim 2$. The procedure of \cite{Cheng:2007xv} can thus be used to
verify the results for $m_N$, $m_X$ and $m_Y$ from the present
procedure and possibly the two can be combined to obtain smaller
errors than from either one, with $m_Z$ determined by the procedure of
this letter.

Overall, we have obtained a highly-encouraging level of accuracy for
the mass determinations in events with two chains terminating in an
invisible particle. Once the masses are known with this level of
accuracy, the next step will be to examine detailed distributions for
various possible models (MSSM, little-Higgs, Universal Extra Dimensions), assuming the
determined masses and keeping only solutions for each event
consistent with them. The different models can be expected to predict
sufficiently distinct distributions (for the same mass choices) that
the precise nature of the invisible particle can be determined. We
will then be able to make fairly precise predictions for its relic density and
check for consistency with observation. Showing that the dark matter
particle as observed at the LHC 
predicts a relic density consistent with cosmological observations
would resolve one of the most important issues of modern-day physics.

\acknowledgments 
This work was supported in part by U.S. Department of Energy grant No. DE-FG03-91ER40674.

\end{document}